# Control of a Josephson Digital Phase Detector via an SFQ-based Flux Bias Driver

**LAURA DI MARINO[1], LUIGI DI PALMA[2], MICHELE RICCIO[1], FRANCESCO FIENGA[1], MARCO ARZEO[2], OLEG MUKHANOV[3]**

[1]University of Naples Federico II, 80125 Naples, Italy (email: michele.riccio@unina.it, francesco.fienga@unina.it)

[2]SEEQC-EU, 80146 Naples, Italy (email: marzeo@seeqc.com, lpalma@seeqc.com)

[3]SEEQC, Inc., 10523 Elmsford NY, USA (email: omukhanov@seeqc.com)

**ABSTRACT** Quantum computation requires high-fidelity qubit readout, preserving the quantum state. In the case of superconducting (SC) qubits, readout is typically performed using a complex analog experimental setup operated at room temperature, which poses significant technological and economic barriers to large system scalability. An alternative approach is to perform a cryogenic on-chip qubit readout based on a Josephson Digital Phase Detector (JDPD): a flux switchable device capable of digitizing the phase sign of a coherent input. The readout operation includes the flux excitation of the JDPD to evolve from a single to a double-minima potential. In this work, the effect of the flux bias characteristics on the JDPD performances is studied numerically. To meet the identified requirements that maximize detection fidelity and tackle the engineering challenges, a cryogenic on-chip Single Flux Quantum based flux bias driver is proposed and discussed.

## I. INTRODUCTION

Fast high-fidelity quantum non demolition (QND) qubit readout is an indispensable requirement in the development of fault tolerant quantum computers [1]–[4]. In fact, a lot of proposed applications, such as error correction [5]–[7], teleportation [8], [9] and state initialization [10], [11], necessitate real-time, measurement-based feedback [12] in which preserving the qubit state during the readout operations is crucial to avoid logical errors.

In the superconducting quantum architecture, the most commonly used technique to probe qubit state is based on measuring the qubit state-dependent dispersive frequency shift of a coupled resonator [1], [13], [14]. The readout resonator can be probed either in reflection or transmission with a tone at a fixed frequency, which will encode the information on the qubit state in the relative phase and amplitude [15]. While in first experiments, averaging was necessary to determine the qubit state with negligible detection error [1], the advent of near quantum-limited noise amplifiers [16]–[18] made it possible to perform single shot readout with fidelity exceeding 99 % [3], [12]. The most advanced readout architectures include also Purcell Filters [19]–[22], which protects qubit from spontaneous emission while allowing for fast probing of the readout resonator with integration time down to 50- 100 ns [3], [12].

However, despite the last advancements, the generation, demodulation and digitization of the readout signal are still done at room temperature (RT), which limits the speed of measurement-based feedback operations and the scalability to a large number of qubits [23]. The main engineering challenges to system scalability stand in the need of physically large microwave components, such as high-bandwidth lines from RT to millikelvin temperatures, non-reciprocal microwave isolators, needed to properly handle the readout tones while protecting the fragile qubit coherence, large latency in room temperature demodulation and digitizing electronics [24]–[26].

These challenges can be addressed with cryogenic superconducting readout electronics deployed in a cryostat in relatively close proximity to the qubits [24], [25], [27], [28].

Josephson Photon Multiplier (JPM) [26], [29], a microwave-frequency analog of the avalanche photodiode, was the first superconducting qubit readout scheme capable of a substantial improvement of the readout chain. However, the JPM still requires its control signals generated by room-temperature electronics.

A different approach, based on a flux-tunable Josephson Digital Phase Detector (JDPD), has been proposed in [13], [30]. By means of external flux pulses, the JDPD can diabatically switch from a single to a double-well potential energy configuration. This feature can be employed to digitize the phase of a microwave coherent tone applied to its input node [30]. The phase response of a readout resonator can be hence detected to infer the information on the qubit state. [13]. Integrated with SFQ-based digital circuitry, the JDPD approach allows for fast and high-fidelity readout and provides a digital output directly at the mK stage. This solves the latency and bandwidth bottlenecks, and reduces the need for bulky microwave components, making the whole readout process more accurate, reliable and energy-efficient [13]. These are all key ingredients for a scalable, fault-tolerant and hence application-relevant quantum computing platform.

A crucial step in the proposed phase detection technique is the evolution of the JDPD potential from a single- to a double-well configuration by applying a diabatic flux bias waveform (flux-switch) [13], [30]. In this work, we investigate how phase detection is influenced by the characteristics of the flux-switch. We perform classical simulation including circuit noise using PSCAN2 [31]. We identify an optimal flux-switch duration range to maximize the phase detection fidelity. We analyze how the JDPD operations are robust with respect to flux noise fluctuations. To achieve the best performances, we propose an SFQ flux driver as possible candidate to provide such flux switching pulse. Since it is based on energy-efficient SFQ logic [32]–[35] this technology is capable of operating at very high speeds (tens of gigahertz clock) and has low power dissipation of the order of few nW. As a result, the circuit could be safely located contiguously to the quantum chips. This can also enable the required phase locking between the input pulse and flux-switch pulse making the detection possible with high fidelities.

## II. THE JOSEPHSON DIGITAL PHASE DETECTOR

The Josephson Digital Phase Detector (JDPD) equivalent circuit is reported in Fig. 1(a): the JDPD consists of two RF-SQUIDs ($I_{c1}$ $L_1$ $L$ and $I_{c2}$ $L_2$ $L$) sharing the inductive load ($L$), similar to a Quantum Flux Parametron (QFP) [36]–[39]. The device can be tuned via magnetic flux $\phi_1$ and $\phi_2$ through the RF-SQUIDs as shown in Fig. 1(a). The single $\phi_{1,2}$ fluxes can be expressed in terms of their linear combination $\phi_{+,-}$, defined as

$$\phi_+ = \frac{\phi_1 + \phi_2}{2}$$
$$\phi_- = \frac{\phi_1 - \phi_2}{2} \qquad (1)$$

The $\phi_+$ flux is the one that flows through the two RF-SQUIDs, while the $\phi_-$ flux unbalances the magnetic flux contribution between them. So, by playing with $\phi_{1,2}$ and by properly stimulating the JDPD, it is possible to generate multiple potential energy profiles, as described in more details in [13], [30]. Cases of particular interest occur when $\phi_+ = \pi$, corresponding to a double well potential configuration, and when $\phi_+ = \pi/2$, for which the potential is full harmonic. The possibility to change rapidly the potential configuration allows the phase detection of a GHz input tone [13], [30].

The protocol is shown in Fig. 1(b) and adheres to the following steps:
- step (i): the JDPD is in its harmonic configuration, with $\phi_+ = \pi/2$;
- step (ii): a current stimulus $I(t)$ is sent to the JDPD input node. This situation can be analyzed from the perspective of a particle with coordinates:

$$\varphi \equiv \frac{2\pi L I_L}{\phi_0} \qquad (2)$$

  where $I_L$ is current in the inductor $L$. Under the drive provided by $I(t)$, the phase particle follows the time evolution of the input signal, transferring the information on its phase to the JDPD;
- step (iii): the information is digitized with a diabatic flux pulse that brings $\phi_+$ to $\pi$, setting the JDPD potential shape to the double well configuration. The phase particle will collapse in one of the two states depending on where it was located when the flux-switch occurred with respect to $\varphi = 0$. These two states are referred to as 0 and 1 and correspond to the phase particle collapsing respectively in the left or right well;
- step (iv): the phase particle position can be sensed by measuring the current flowing through the JDPD central inductor, since the left and right states are associated to opposite value for $I_L$ (+ right loop, − left loop).

A crucial aspect of the phase detection process is the flux switch $\phi_+ = \pi/2 \to \pi$, which is the key operation to achieve high fidelity readout. Intuitively, the flux switch should have faster rise time compared to the input tone period, to avoid unwanted phase particle oscillation during the emergence of the barrier between the two wells. At the same time, higher slew rate flux switches provide more energy to the system giving the possibility to the phase particle to escape from the trapping well.

To ensure measurements reproducibility, a precise phase-locking between the stimulus tone and the flux switch is also required. The value of $\phi_+$ should have minimal noise fluctuations to maximize the detection fidelity. In the next sections, we analyze all these aspects with numerical simulations. Starting from the JDPD characteristics of [13], [30], we identify the optimal working parameters for the flux switch to ensure the best performances possible.

We also introduce an SFQ-based Flux Bias Driver to meet the desired specifications in terms of flux noise, speed operations and phase locking. Based on the SFQ logic, this device can be also assisted by an SFQ based processor located inside the fridge, bringing benefits to reduce the number of physical connections to the room temperature setup.

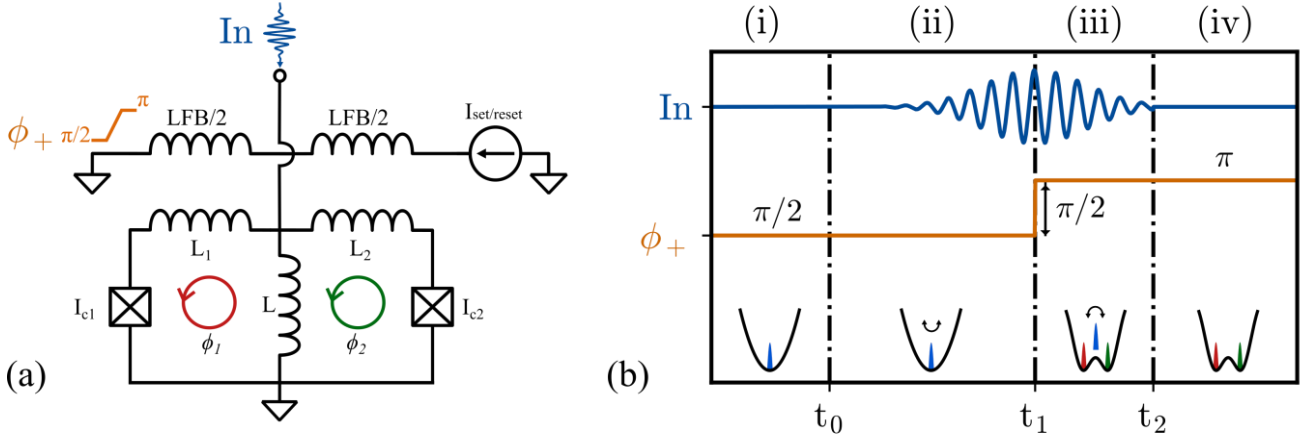

**FIGURE 1.** (a) The Josephson Digital Phase Detector (JDPD) Circuit. It is composed of two RF-SQUIDs sharing the central inductor load L. The current generator $I_{set/reset}$ with the inductances *LFB*/2 schematize flux switch generation; the blue tone is the stimulus signal sent to the input node of the detector. (b) Phase detection protocol, adapted from [13]: (i) Ready state, the potential is in an harmonic configuration (single well) as $\phi_+ = \pi/2$; (ii) Detection state, an input stimulus is sent to JDPD, which causes oscillations of the phase particle; (iii) Digitalization state, the flux achieves $\pi$ and the potential goes to a double-minima configuration; (iv) Sense state, the stimulus is turned off and eventually the phase particle is collapsed in either of the wells.

## III. SIMULATIONS WITH NOISE

The JDPD behavior and phase detection protocol described in Fig. 1 have been simulated within PSCAN2 software [31]. The flux drive is reproduced by using two phase generators placed between $L_{1,2}$ and $J_{1,2}$ which provide the $\phi_{1,2}$ fluxes. We consider the JDPD with $L = 200$ pH, $I_c = 6$ μA and $L_1 = L_2 = 20$ pH, as in the real device reported in [13],[30]. The Josephson junctions have been modelled within the framework of the Resistevely and Capacitively Shunted Junction (RCSJ) model [40]. This model considers a Josephson junction as a parallel combination of a superconducting current channel, a normal resistive current path, and a capacitive element and it captures the interplay between these components, providing insights into the junction's dynamic response to external stimuli [40], [41]. The RCSJ model of a Josephson junction can be analogized to the model of a simple pendulum. According to this analogy, the phase difference $\phi$ across the Josephson junction corresponds to the angle $\theta$ of the pendulum. The superconducting current term $Ic \sin(\phi)$ is analogous to the gravitational torque acting on the pendulum, which is proportional to $\sin(\theta)$. This represents the restoring force that brings the pendulum back to its equilibrium position. The resistive component $R$ represents a damping force that is proportional to the angular velocity of the pendulum, similar to friction in the pendulum system. The junction damping plays an important role when simulating the JDPD behavior, as it gives a measure about the phase particle oscillations magnitude. In fact, it should not be too small in order to prevent large oscillations of the phase particle, that would affect the detection results. In particular, for smaller values of $R$ there will be small phase particle oscillations and viceversa. Lastly, the capacitance $C$ represents the inertia of the pendulum. In the mechanical system, this is analogous to the moment of inertia of the pendulum, which determines how the pendulum responds to torques. According to the fabrication process of the Josephson junctions [42], the Stewart-McCumber parameter $\beta_C$ defined as:

$$\beta_C = \frac{2\pi V^2 C_S}{\phi_0 J_S} \quad (3)$$

where $C_S$ is the capacitance per unit area of $50\ fF/\mu m^2$, $J_S$ is the critical current density of $10\ \mu A/\mu m^2$ and $V_c = I_C R$, where $R$ is the intrinsic junction resistance of $310\ \Omega$, results in value of $\beta_C = 53$.

The input stimulus (shown in Fig. 1(a)), which eventually could be the readout tone from the qubit resonator, is generated with a current source attached to the JDPD's input node. The stimulus consists of a sinusoidal pulse, modulated with a Gaussian envelope, with a frequency of 7.5 GHz. This value is typical for readout resonators used in circuit quantum electrodynamics (cQED) [15], [22]. The input stimulus amplitude has been optimally chosen to excite the phase particle without provoking large oscillations that would alter the phase detection result. The stimulus duration is calculated depending on the flux-switch time length, as it is a parameter that we have evaluated in our simulations. Thence, we adjust the stimulus duration such that flux-switch always happens at half time of the input stimulus. To simulate the device fidelity and how it is affected by the flip duration, we introduce a thermal noise model with current noise generators in parallel to the Josephson junctions following the method in [43]. The magnitude of the current noise is a Gaussian distributed random number with standard deviation given by:

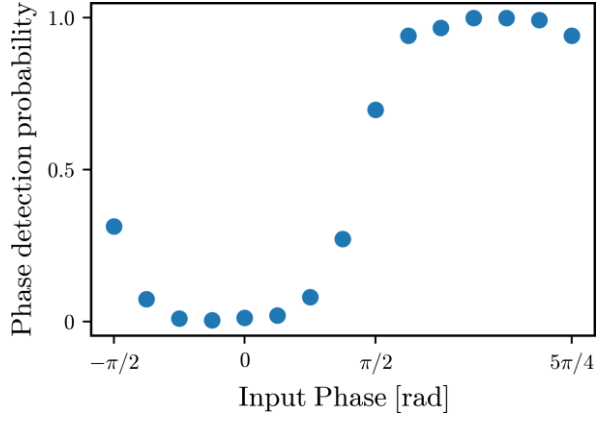

**FIGURE 2.** Phase detection probability versus input stimulus phase, for Josephson junction with $I_C = 6\mu A$.

$$\sigma = \sqrt{\frac{2k_B T}{R\delta t}} \quad (4)$$

where $k_B$ is the Boltzmann constant, $T$ is the noise temperature, $R$ is the junctions normal resistance, and $\delta t$ is the simulation time step set to be 1 ps in our numerical analysis. According to Eq. 4, the noise bandwidth $\Delta f = 2/\delta t$, clearly depends on the simulation time step, meaning that more accurate analysis will have larger noise fluctuations. Hence, we adopt a low pass filter with cut-off frequency equal to the plasma frequency:

$$\omega_P = \sqrt{\frac{2\pi J_S}{\phi_0 C_S}} \quad (5)$$

With this approach, the noise bandwidth [43] is restricted to the characteristic oscillating frequency of the Josephson junctions and the dependence with $\delta t$ is removed.

The JDPD is designed to work in close proximity to the qubit chip; therefore a natural choice for the value T in Eq. 4 would be 10 mK. However, at these temperatures, quantum fluctuations are expected to be much larger than thermal ones [44] and there is the risk of overestimating the detection fidelity. We consider as noise temperature, the crossing temperature between the quantum and classical regime $T_{cross}$ defined as:

$$T_{cross} = \sqrt{\frac{\hbar \omega_P}{2\pi k_B}} \quad (6)$$

$T_{cross}$ stands as an upper limit for noise scale; therefore, within this approach, the simulated performances of our device are expected to be worse than the ones in a real device. By introducing the intrinsic noise of the Josephson junctions, the simulation is repeated 500 times to build statistical distributions. We count for each iteration how many times the phase particle collapses in either the two wells and we extract the phase detection probability. An example of simulation outcome is depicted in Fig. 2, where the phase detection probability is plotted as a function of the input stimulus phase.

The most used parameter to check on the detection quality of this type of devices is the Gray Zone ($\Delta\phi_x$) [45], extracted as a result of fitting the detection probability by the following expression:

$$P = \frac{1}{2}(1 + erf(\sqrt{\pi}\frac{\phi_x - \phi_t}{\Delta t})) \quad (7)$$

where $\phi_x$ is the input tone phase displacement and $\phi_t$ is a fit parameter indicating where the probability is 0.5. Even though utilizing the Gray Zone is a very diffused metric, if the detection probability does not reach its maximum value of 1, the fit quality degrades and the Gray Zone cannot be properly estimated. In thoose situations, we switch to a different metric and evaluate the JDPD separation fidelity , defined as the average between the maximum probability and one minus the minimum probability. Therefore, with the term "separation fidelity" we refer to a parameter that ranges between 0.5 and 1, angd suggests how good the phase detection is: values close to 1 stand for a precise discrimination between state 0 and 1. In the following sections, the effects of the generated flux switch on the phase detection will be discussed in terms of phase detection probability, separation fidelity and gray zone results.

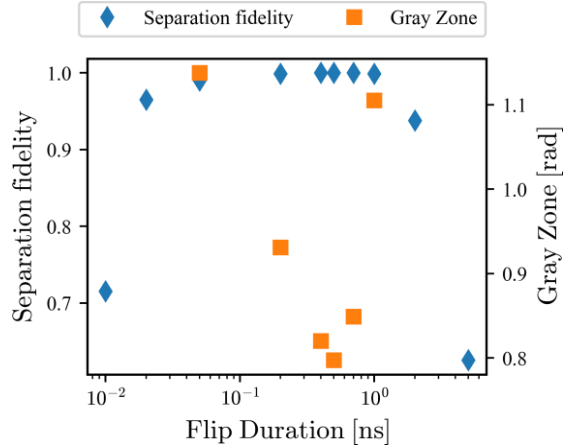

**FIGURE 3.** Separation fidelity (blue diamonds) and Gray Zone (orange squares) versus flip duration for $I_C = 6\,\mu A$: both slower and faster flip rise time result in a degradation of the phase detection probability, while for duration in the range of nanoseconds (100 ps − 1 ns), the separation fidelity (gray zone) achieves its maximum (minimum) value.

## IV. FLUX FLIP DURATION EFFECTS ON PHASE DETECTION

The flux flip duration must be carefully considered when evaluating the JDPD separation fidelity. The slope of the flux-switch strongly influences the quality of the detection. For this purpose, numerical simulations with a sweep on the flux-switch duration has been carried out in a range from 10 ps to 5 ns. For each value of duration, the detection probability is estimated over 15 values of the input tone phase in the range $[0, 2\pi]$, with over 500 repetitions per phase value. Fig. 3 displays the obtained results for JDPD separation fidelity (blue diamonds) and the Gray Zone (orange squares) as a function of the flux switch duration. It is important to highlight that in all the simulations presented in this work, an upper threshold of $\pi/2$ is set for the Gray Zone, as values above this limit correspond to an adverse fitting of the function in Eq. 7. From Fig. 3, it is possible to highlight three different regions depending on the duration of the flip:

- region for flip durations up to 50 ps, where the separation fidelity reaches a minimum of 0.7. This is due to the fact that the flip duration is comparable with the input tone period, and suggests that the states' discrimination is not accurate;
- region for flip duration above 1 ns, with the separation fidelity reaching its lowest value of 0.6. In this circumstance, a greater energy is given to the system, which makes the phase particle escape the well it was trapped in;
- optimal region, for flip duration in the range $[100\,ps - 1\,ns]$. The JDPD separation fidelity achieves its maximum value of 1, meaning that the detector is reliably discriminating the probability of being in state 0 or 1. In this situation, the Gray Zone parameter has been evaluated and, from the plot, it is noticeable that the Gray Zone optimal range is even narrower than the separation fidelity one.

Therefore we deduce that the best performances are accomplished for duration ranging from 100 ps to 1 ns.

## V. FLUX FLUCTUATIONS EFFECTS ON PHASE DETECTION

The phase detection protocol key operation is based on the JDPD switch from a single to a double-minima potential configuration. As described in the JDPD section, this transition happens when the flux is set to $\pi$. Realistically, there are flux fluctuations of $\phi_+$ that could be detrimental to detection fidelity. In this section, we discuss numerical simulations in which, in addition to the Junctions intrinsic fluctuations, we add Gaussian white noise to $\phi_+$. The standard deviation is gradually increased up to the maximum value of 0.57 rad. This approach enables us to evaluate how well the system maintains its performance against the flux fluctuations. For those simulations, we consider a flux-switch rise time of 100 ps and a 0.35 µA amplitude for the input stimulus.

Fig. 4(a) illustrates a plot with phase detection probability on the y-axis and input phase varying along the x-axis, while the colorbar on the right displays the considered flux noise range. It is possible to observe that for noise up to 0.29 rad the detection is unaffected with a detection probability 0.99, but for larger fluctuations the JDPD performance starts degrading. This indicates that the device is robust against a 10% fluctuations on the desired $\phi_+$ flux value. The very same information is shown in Fig. 4(b), where on the y-axis are displayed the JDPD separation fidelity (blue diamonds) and Gray Zone (orange squares) and on the x- axis the flux noise. As the noise becomes larger, the detector separation fidelity is considerably reduced and Gray Zone increased.

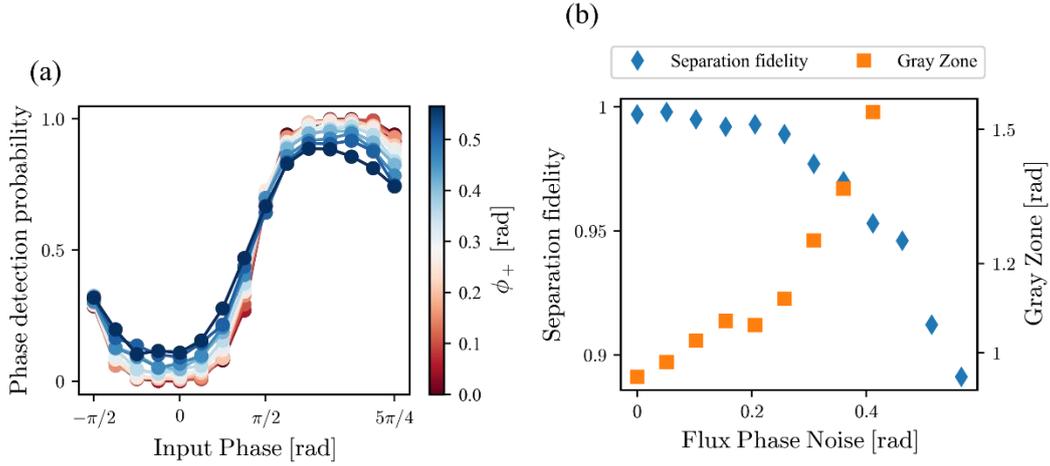

**FIGURE 4.** (a) Phase detection probability versus input phase by adding flux fluctuation to the $\phi_+$ flux flip. For noise up to 0.29 radiant, the system is robust against flux fluctuations, since probability is still around its maximum value. (b) Plot of the JDPD separation fidelity and Gray Zone, of the JDPD: as expected, for small flux noise, the separation fidelity achieves 1, but starts degrading when noise becomes larger, and vice-versa for the Gray Zone.

## VI. FLUX BIAS DRIVER

In previous sections specific requirements for the flux switch, in terms of speed operation and flux noise, have been outlined. To adhere to those specifications, one possible approach is to integrate the JDPD with an SFQ circutry, as it is located adjacent to the JDPD and qubit chips [42]. The proposed SFQ circuit is the adaptation of the Flux Bias Driver (FBD) proposed in [32] whose block diagram can be found in Fig. 5. With the FBD, the generation of the flux switch is moved to cryogenic temperatures, enabling easy control and synchronization with the readout tone, which further improves measurements fidelity. Also, by being placed in the cryostate in close proximity to the JDPD, all the noise and crosstalk introduced from cables needed when generating the flux flip at room temperature get entirely avoided.

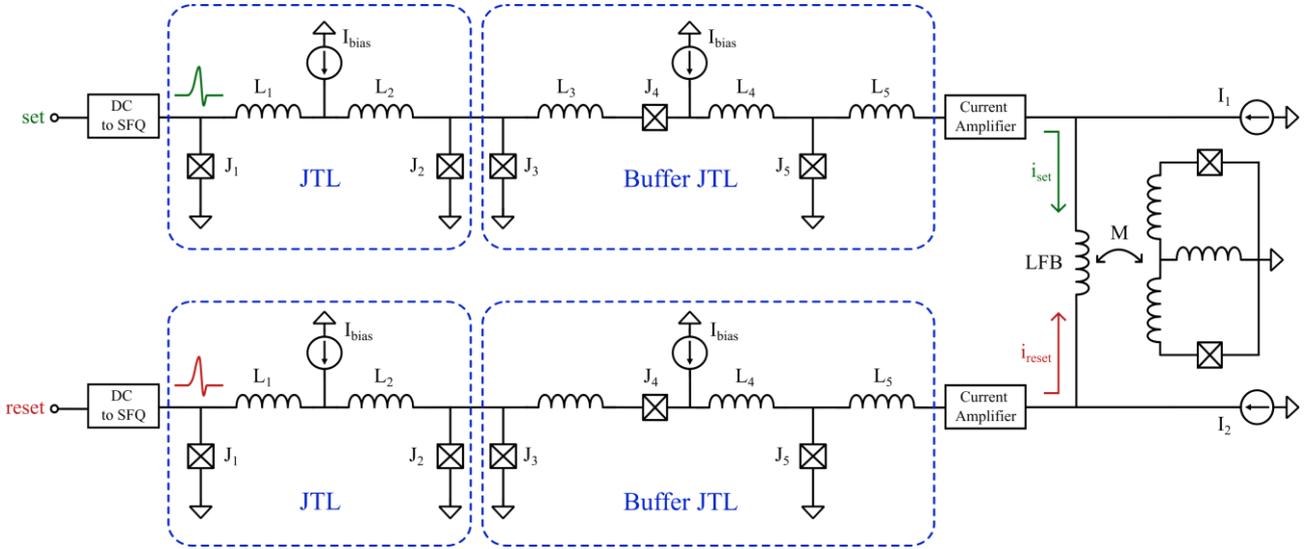

**FIGURE 5.** Flux Bias Driver coupled to Josephson Digital Phase Detector schematic. The set and reset branches are used to drive flux to the JDPD, which is possible thanks to the inductive coupling between the LFB inductor of the FBD and the $L_1$ and $L_2$ inductors of the JDPD. In particular, the set branch sends current until $\phi_+ = M * I_{LFB} = 2\pi$, then the reset branch resets the system to its initial condition.

The FBD consists of two branches to *set* and *reset* the JDPD potential from a single to a double well potential. An inductive coupling between the JDPD and the FBD, through the LFB inductor (Fig. 5), connects the drive flux to the JDPD itself. By sending a current signal through the LFB inductor, the generated $\phi_+$ flux is obtained as:

$$\phi_+ = M * I_{LFB} \qquad (8)$$

where $M$ is the mutual inductance, set to 5.4 pH in the presented results to achieve $\phi_+ = 2\pi$. The behavior of the flux bias driver is the following (Fig. 5):
- A train of triggering pulses is sent to the set node, where a DC to SFQ circuit converts them into SFQ pulses;
- Those SFQ pulses go through a Josephson Transmission Line (JTL) and a Buffer JTL, which ensure that the pulses will propagate in one direction only and will not go back to the set/reset node;

- Then, the Buffer JTL output current gets amplified via an SFQ-based amplifier, before reaching the LFB inductor;
- According to Eq. 8, once $\phi_+ = 2\pi$, the reset branch will bring the whole device to its initial state.

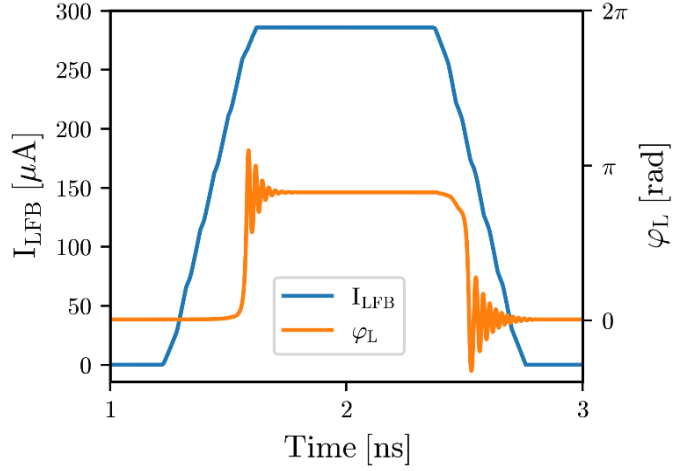

**FIGURE 6.** Flux Bias Driver behavior: the blue line is the current signal flowing through the LFB inductance, orange one is the phase at the ends of the JDPD central inductance. A significant point about this plot is that the set and reset operations are carried out in a range of a few nanoseconds.

Fig. 6 shows the Flux Bias Driver behavior simulated with PSCAN2 simulator. The blue line represents the current on the LFB inductor, whereas the orange one represents the phase at the ends of the of the JDPD's central inductor. Simulation results indicate that the system is properly working with the expected JDPD response. In fact, for $I_{LFB} = 150\ \mu A$ the flux is equal to $\pi$, and the phase at the ends of the $L$ inductor goes from a minimal value to almost $\pi$. This means that the phase particle, after some initial oscillations determined by the damping, has been successfully trapped in one of the two potential wells. Another significant advantage of using an SFQ-based FBD for the JDPD flux control, relies on the possibility of performing the whole set and reset flux driving procedure in a time scale of nanoseconds.

## VII. $I_{LFB}$ SHAPE EFFECTS ON PHASE DETECTION PROBABILITY

Once the duration and noise requirements of the flux flip have been determined and satisfied with the FBD, we now focus on the $I_{LFB}$ current shape and its influence on system performance. In all the previous simulations, a single-step signal has been considered for the set and reset current. We still employ the same waveform for $I_{LFB}$, but now the simulation exploits multiple signal shapes with an increasing number of steps from 1 to 8.

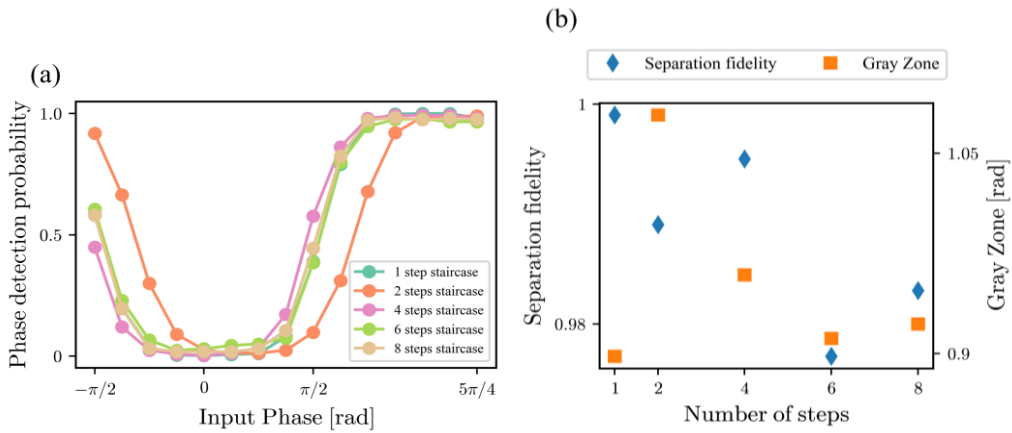

**FIGURE 7.** (a) JDPD separation sensibility vs input phase by varying the shape of the current sent through the LFB inductor. The shape change is in the number of step with which the $\phi_+ = 2\pi$ is achieved. The results suggest that the states' discrimination is accurate for all shapes considered. (b) Gray Zone and separation fidelity versus increasing number of steps of the step signal.

The purpose is to evaluate whether it is better to have just one step that directly reaches $2\pi$, or multiple smaller steps to gradually achieve this value. The results in terms of "phase detection probability" are presented in the plot in Fig. 7(a) which suggests that the phase discrimination between state 0 and 1 is efficient and accurate for almost all shapes considered. The orange curve, the one referring to the 2 steps staircase, appears to be dephased with respect to the others.

This happens as there is a time interval, between the first and second step, in which the step signal is flat before $\phi_+$ reaches $2\pi$. It is important to point out that the phase displacement occurs for all the investigated shapes, but is more evident for the one with two steps. Fig. 7(b) shows a plot with the JDPD separation fidelity (blue diamonds) and Gray Zone (orange squares) on y-axis, and the increasing number of steps over the x-axis. Both separation fidelity and Gray Zone values confirm that the phase detection is very precise as their values are all almost above respectively 0.98 and 0.9, which is synonym of high fidelity and accuracy of the detection protocol. So, we conclude that staircases with single or multiple steps do not have a significant detrimental impact on the phase detection.

### VIII. CONCLUSION

In this work, we have studied how the flux switch influences the JDPD performances. To do that, we have performed noisy simulations using the software PSCAN2. We have estimated an optimal flip duration range in order to achieve the best fidelity possible. Also, we have studied how the flux noise affects the phase detection operation. From this analysis, we have proposed an SFQ flux driver as the optimal circuit for integration with the JDPD. This circuit could be safely located contiguously to the quantum chips and it offers the possibility to operate at the required clock speed and low flux noise. Furthermore, it can ensure the required phase locking between the input and flip pulse making the detection possible with high fidelities.